\documentclass[aps,prd,preprint,eqsecnum,nofootinbib,groupedaddress,showpacs,floatfix]{revtex4}

\usepackage[dvips]{graphicx}
\usepackage{multirow}
\newcommand{\BlackHat}{{\sc BlackHat}}

\newcommand{\SHERPA}{{\sc SHERPA}}

\newcommand{\AMEGIC}{{\sc AMEGIC++}}
\newcommand{\COMIX}{{\sc COMIX}}
\newcommand{\SISCone}{{\sc SISCone}}
\newcommand{\ntuple}{{$n$-tuple}}
\newcommand{\ntuples}{{$n$-tuples}}

\newif\ifdraft
\draftfalse
\newif\ifpreprint
\preprinttrue

\def\fig#1{fig.~{\ref{#1}}}

\def\figs#1#2{figs.~{\ref{#1}} and {\ref{#2}}}

\def\sect#1{section~{\ref{#1}}}
\def\eqn#1{eq.~(\ref{#1})}

\def\eqns#1#2{eqs.~(\ref{#1}) and~(\ref{#2})}

\def\tab#1{table~{\ref{#1}}}
\def\Tab#1{Table~{\ref{#1}}}

\def\nn{\nonumber}
\def\qb{\bar q}

\def\Qb{\bar Q}

\def\eps{\epsilon}
\def\nub{\bar \nu}

\def\Wj{$W\,\!+\,1$}
\def\Wjj{$W\,\!+\,2$}
\def\Wjjj{$W\,\!+\,3$}
\def\Wjjjj{$W\,\!+\,4$}
\def\Wjjjjj{$W\,\!+\,5$}
\def\Wjjjjx{$W\,\!+\,4,5$}

\def\Wmjjjjj{$W^-\,\!+\,5$}

\def\Wpjjjjj{$W^+\,\!+\,5$}

\def\WZjjj{$W,Z\,\!+\,3$}

\def\Wjn{$W\,\!+\,n$}
\def\Wjnp1{$W\,\!+\,(n\!+\!1)$}
\def\Wmjn{$W^-\,\!+\,n$}

\def\Wpjn{$W^+\,\!+\,n$}

\def\Wpmjn{$W^{\pm}\,\!+\,n$}
\def\Wpmjnm{$W^{\pm}\,\!+\,(n\!-\!1)$}
\def\Wjnm{$W\!+\,(n\!-\!1)$}

\def\Zjn{$Z\,\!+\,n$}

\def\WZjjj{$W,Z/\gamma^*\,\!+\,3$}
\def\WZjjjj{$W,Z/\gamma^*\,\!+\,4$}
\def\gjn{$\gamma\,\!+\,n$}
\def\jet{{\rm jet}}
\def\pT{p_{\rm T}}

\def\root{{\sc root}}
\def\Ord{{\cal O}}

\def\ETsl{{\s E}_{\rm T}}

\def\HTpartonic{{\hat H}_{\rm T}}
\def\HTpartonicp{{\hat H}_{\rm T}'}

\def\HTjets{H_{\rm T}^{\rm jets}}

\newbox\charbox
\newbox\slabox
\def\s#1{{      
        \setbox\charbox=\hbox{$#1$}
        \setbox\slabox=\hbox{$/$}
        \dimen\charbox=\ht\slabox
        \advance\dimen\charbox by -\dp\slabox
        \advance\dimen\charbox by -\ht\charbox
        \advance\dimen\charbox by \dp\charbox
        \divide\dimen\charbox by 2
        \raise-\dimen\charbox\hbox to \wd\charbox{\hss/\hss}
        \llap{$#1$}
}}

\begin{document}

\title{
\ifpreprint
\hbox{\rm\small
UCLA/13/TEP/101 $\null\hskip 3.3 cm \null$
SB/F/420-13$\null\hskip 4.0 cm \null$
SLAC-PUB-15395}
\hbox{\rm\small
IPPP-13-16  $\null\hskip 3.3 cm \null$ \hfill
 $\null\hskip 7.1 cm \null$
FR-PHENO-2013-004}
\hbox{$\null$\break}
\fi
Next-to-Leading Order \Wjjjjj-Jet Production at the LHC}

\author{Z.~Bern${}^a$, L.~J.~Dixon${}^b$, F.~Febres Cordero${}^c$, S.~H{\" o}che${}^b$, H.~Ita${}^{d}$, D.~A.~Kosower${}^{e}$, D.~Ma\^{\i}tre${}^{f}$ and K.~J.~Ozeren${}^a$
\\
$\null$
\\
${}^a$Department of Physics and Astronomy, UCLA, Los Angeles, CA
90095-1547, USA \\
${}^b$SLAC National Accelerator Laboratory, Stanford University,
             Stanford, CA 94309, USA \\
${}^c$Departamento de F\'{\i}sica, Universidad Sim\'on Bol\'{\i}var, 
 Caracas 1080A, Venezuela\\
${}^d${Physikalisches Institut, Albert-Ludwigs-Universit\"at Freiburg,
       D--79104 Freiburg, Germany}\\
${}^e$Institut de Physique Th\'eorique, CEA--Saclay,
          F--91191 Gif-sur-Yvette cedex, France\\
${}^f$Department of Physics, University of Durham, Durham DH1 3LE, UK\\
}

\begin{abstract}
We present next-to-leading order QCD predictions for the total cross
section and for a comprehensive set of transverse-momentum distributions
in \Wjjjjj-jet production at the Large Hadron Collider.
We neglect the small contributions from subleading-color virtual terms, 
top quarks and some terms containing four quark pairs.
We also present ratios of total
cross sections, and use them to obtain an extrapolation formula
to an even larger number of jets.
We include the decay of the $W$ boson into
leptons.  This is the first such computation with six final-state
vector bosons or jets.  
We use \BlackHat\ together with \SHERPA\ to carry out the computation.
\end{abstract}

\pacs{12.38.-t, 12.38.Bx, 13.87.-a, 14.70.Fm \hspace{1cm}}

\maketitle

\section{Introduction}

Reliable theoretical predictions for Standard-Model processes at the
Large Hadron Collider (LHC) are
important to ongoing searches for new physics.  They are also
important to the increasingly precise studies of the newly discovered
Higgs-like boson~\cite{AtlasHiggs,CMSHiggs}, of the top quark,
and of vector boson self-interactions.  New-physics
signals very typically lie beneath Standard-Model backgrounds in a
broad range of search strategies.  Ferreting out the signals requires
a good quantitative understanding of the backgrounds and their
uncertainties.  With the increasing jet multiplicities used
in cutting-edge search strategies, this becomes more and more
challenging.  Some of the uncertainty surrounding predictions of
Standard-Model background rates can be alleviated through use of
data-driven estimates, but this technique also requires theoretical
input to predict the ratios of background processes in signal regions
to those for control processes or in control regions.

Predictions for background rates at the LHC
rely on perturbative QCD, which enters all aspects of
short-distance collisions at a hadron collider.  Leading-order (LO)
predictions in QCD suffer from a strong dependence on the unphysical
renormalization and factorization scales.  This dependence gets
stronger with increasing jet multiplicity.  Next-to-leading (NLO)
results generally reduce this dependence dramatically, typically to a
10-15\% residual sensitivity.  Thus they offer the first quantitatively
reliable order in perturbation theory.

The production of a $W$ boson in association with jets has played a special
role in collider physics.  It was the dominant background to top-quark
pair production at the Tevatron. At the LHC it remains an important
background for precision studies, including those of top quarks.  It
is important to many new physics searches involving missing energy,
including those for supersymmetry.  Recent searches have made use of
samples with high jet multiplicity, and proposed searches aim to push
to higher multiplicities yet.   Precise quantitative control over the
theoretical predictions leads to improved sensitivity to new phenomena.
Measurements of $W$ boson production
in association with multiple jets have been made by the
CDF~\cite{CDFWjets} and D0~\cite{D0Wjets} collaborations at the
Tevatron, and by the ATLAS~\cite{ATLASWjets,NTupleUse} and
CMS~\cite{CMSWjets} collaborations at the LHC.
Such measurements also permit
stringent tests of the predictions of the Standard Model.  

Theoretical predictions for the production of vector bosons with a
lower multiplicity of jets (one or two jets) have been available at
NLO in QCD for many years~\cite{NLOW1jet,MCFM,FernandoWjetpapers}.  In
recent years, the advent of new on-shell
techniques~\cite{UnitarityMethod, DDimUnitarity, Zqqgg, NewUnitarity,
  BCFW, GenHel, OPP, Forde, Badger} for computing one-loop amplitudes
at larger multiplicity has led to NLO results for
three~\cite{W3jPRL,EMZW3j,W3jDistributions,BlackHatZ3jet} and
four~\cite{W4j,Z4j} associated jets.  Other new results include those
for the production of vector-boson pairs~\cite{NLOVVjj} or 
top--anti-top pairs~\cite{BDDP,NLOttjj}
in association with two jets.  Another recent approach~\cite{Weinzierl}
has been demonstrated in the production of up to seven jets in $e^+e^-$
collisions, and shows promise for LHC physics as well.  There have also
been important advances with more traditional methods, especially for
the case of heavy quarks~\cite{HeavyQuarkFeynman,BDDP,OpenLoops}.  In
the present article, we take another step forward in multiplicity,
presenting NLO results for inclusive \Wjjjjj-jet production at the
LHC.  These are the first NLO QCD results at a hadron collider with
six or more electroweak bosons or jets in the final state.  We
incorporate the decay of the $W$ boson into leptons, so that there are
seven final-state objects to track.

In the present paper we use on-shell methods as implemented
in numerical form in the \BlackHat{} software library~\cite{BlackHatI}.
This library, together with the \SHERPA{}
package~\cite{Sherpa}, has previously been used to make NLO
predictions for \WZjjj-jet
production~\cite{W3jDistributions,BlackHatZ3jet}, for \WZjjjj-jet
production~\cite{W4j,Z4j}, and for four-jet
production~\cite{FourJets}.  It has also been used in investigations
of high-$\pT$ $W$ polarization~\cite{Wpolarization}, and to
compute \gjn-jet to \Zjn-jet ratios for assessing
theoretical uncertainties~\cite{PhotonZ,PhotonZ3} in the CMS
searches~\cite{PhotonZexp} for supersymmetric particles.  
The ATLAS collaboration has also used results from
\BlackHat{} computations with \SHERPA{} for Standard-Model studies of
electroweak vector-boson production in association with three or more
jets~\cite{NTupleUse}.  Other programs that use on-shell methods
are described in refs.~\cite{OtherOnShellPrograms}.

\SHERPA{} is used to manage the numerous partonic subprocesses
entering the calculation, to integrate over phase space, to construct
physical distributions, and to output \root{}~\cite{ROOT} $n$-tuples.
In contrast to earlier computations, we use the \COMIX{}
package~\cite{Comix} to compute Born and real-emission matrix
elements, along with the Catani--Seymour~\cite{CS} dipole subtraction
terms.  Rather than repeating the entire computation for each scale
and for each parton distribution function (PDF) set, we store
intermediate results in $n$-tuple format, recording momenta for all
partons in an event, along with the coefficients of various scale- or
PDF-dependent functions in the event weight.  The $n$-tuple storage
makes it possible to evaluate cross sections and distributions for
different scales and PDF error sets.  We perform the basic calculation
with loose cuts, also making it possible to choose different
(tighter) cuts without recomputing the time-consuming matrix elements.

In this paper, we compute the total cross sections at NLO for
inclusive \Wpjn-jet and \Wmjn-jet production with $n\le 5$ and
describe $W^+/W^-$ ratios and \Wjn-jet/\Wjnm-jet ratios.  Such ratios
can be sensitive probes of new physics.  We also study two types of
distributions: the differential cross section in the total hadronic
transverse energy $\HTjets=\sum_{j\in {\rm jets}} p_{\rm T}^j$, and
the complete set of differential cross sections in the jet transverse
momenta.  For four and five jets we make use of a leading-color
approximation for the virtual contributions.  This approximation has
been shown to have subleading-color corrections of under 3\% for
processes with four or fewer associated
jets~\cite{W3jDistributions,ItaOzeren}.

This paper is organized as follows.  In \sect{BasicSetUpSection} we 
summarize the basic setup of the computation.  In \sect{ResultsSection}
we present our results for cross sections, ratios and distributions. 
We give our summary and conclusions in \sect{ConclusionSection}.

\section{Basic Setup}
\label{BasicSetUpSection}

\begin{figure}[t]
\includegraphics[clip,scale=0.46]{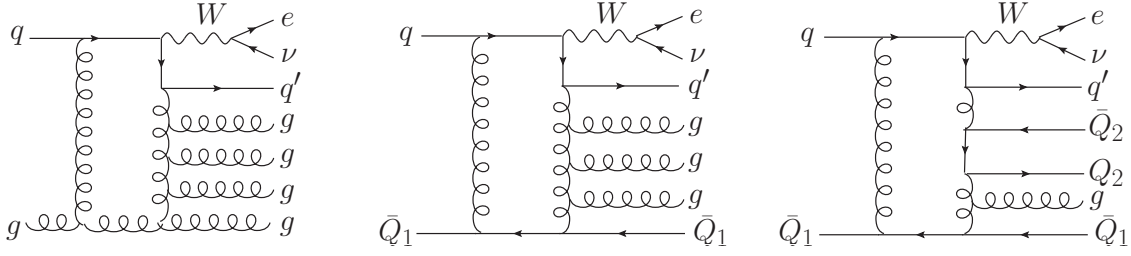}
\caption{Sample eight-point loop diagrams for the processes
$q g \rightarrow W q' \! g g g g$,
$q \bar Q_1 \rightarrow W  q' ggg \bar Q_1$
and $q \bar Q_1 \rightarrow W  q' \bar Q_2 Q_2 g  \bar Q_1$,
followed by the decay of the $W$ boson to leptons. }
\label{lcdiagramsFigure}
\end{figure}

\begin{figure}[t]
\includegraphics[clip,scale=0.52]{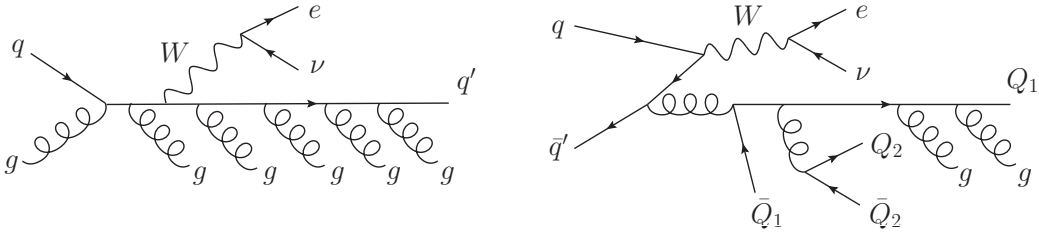}
\caption{Sample nine-point real-emission diagrams for the
processes $q g \rightarrow W  q' ggggg$ and 
  $q \bar q' \rightarrow W \, Q_1 g g Q_2 \bar Q_2 \bar Q_1$,
 followed by the decay of the $W$ boson to leptons.
}
\label{TreeDiagramsFigure}
\end{figure}

In this paper we compute the \Wjjjjj-jet processes in NLO QCD, 
followed by leptonic $W$-boson decay,
\begin{eqnarray}
&& pp \, \longrightarrow
W^{-}+5\, \hbox{ jets} \rightarrow e^{-}\bar \nu_{e}+5 \hbox{ jets}\,,\nn \\
&& pp \, \longrightarrow
W^{+}+5\, \hbox{ jets} \rightarrow e^{+} \nu_{e} + 5 \hbox{ jets}\,.
\end{eqnarray}
These processes receive contributions from several partonic subprocesses.
At leading order, and in the virtual NLO contributions,
the $W^-$ subprocesses are all obtained from
\begin{eqnarray}
 && q \qb' ggggg\rightarrow W^{-}\rightarrow
  e^{-}\, \nub_e \,, \nn \\
 && q \qb' Q_1 \Qb_1 g g g\rightarrow W^{-}\rightarrow
  e^{-}\, \nub_e \,,\nn \\
 && q \qb' Q_1 \Qb_1 Q_2 \Qb_2 g\rightarrow W^{-}\rightarrow
  e^{-}\, \nub_e \,,
\label{Wm5subprocesses}
\end{eqnarray}
by crossing five of the partons into the final state. 
Similarly, we obtain the subprocesses for the $W^+$ case from the
various crossings of the subprocesses
\begin{eqnarray}
 && q \qb' ggggg\rightarrow W^{+}\rightarrow
  e^{+}\, \nu_e \,, \nn \\
 && q \qb' Q_1 \Qb_1 g g g\rightarrow W^{+}\rightarrow
  e^{+}\, \nu_e \,,\nn \\
 && q \qb' Q_1 \Qb_1 Q_2 \Qb_2 g\rightarrow W^{+}\rightarrow
  e^{+}\, \nu_e \,.
\label{Wp5subprocesses}
\end{eqnarray}
The $W$ boson changes the quark flavor and couples to the $q$-$q'$
line. Both the labels $q$ and $Q_i$ denote light quarks. Amplitudes
with multiple identical quark flavors are obtained by appropriate
symmetrization.  Sample Feynman diagrams illustrating virtual
contributions with 1, 2 and 3 external quark pairs are shown in
\fig{lcdiagramsFigure} (although our calculation is not based
on Feynman diagrams).  All contributions to the virtual corrections
are included in a leading-color approximation described below. Besides
the virtual contributions, NLO QCD requires also real-emission
contributions with an additional parton in the final state.  Here we
keep the full color dependence.  However, we drop the finite
contributions from tree amplitudes with four external quark pairs;
they contribute well under 1\% to the cross section.  Sample
real-emission diagrams are displayed in \fig{TreeDiagramsFigure}.

The decay of the vector boson ($W^{\pm}$) into a charged lepton and
neutrino is included at the amplitude level; no on-shell
approximation is made for the $W$ boson.  The lepton-pair invariant
mass follows a relativistic Breit-Wigner distribution with width given
by $\Gamma_W = 2.06$~GeV and mass $M_W = 80.419$~GeV.
(The other electroweak parameters are also chosen as in
ref.~\cite{W3jDistributions}.)  We take the leptonic decay products to
be massless.  In this approximation, of course, the results for muon
final states are identical to those for electrons.  The five
light quarks, $u,d,c,s,b$, are all treated as massless.  We do not
include contributions to the amplitudes from a real or virtual top
quark; its omission should have a percent-level effect on the overall
result~\cite{W4j,Z4j}.  We also approximate the Cabibbo-Kobayashi-Maskawa
matrix by the unit matrix.  As previously determined for the three-jet
case, this approximation causes a change of under 1\% in total cross
sections for the cuts we impose~\cite{W3jDistributions}, and should also
be completely negligible in our study.

\subsection{Kinematics and Observables}

We use standard kinematic variables, whose definitions may be
found in Appendix~A of ref.~\cite{BlackHatZ3jet}.
The renormalization
and factorization scales in ref.~\cite{W3jDistributions}
were chosen as multiples of a total partonic
transverse energy $\HTpartonic$.  We will use a modified version
of it here,
\begin{equation}
\HTpartonicp \equiv \sum_m p_{\rm T}^m + E_{\rm T}^W\,,
\label{HTp}
\end{equation}
where the sum runs over all final-state partons $m$ and 
$E_{\rm T}^W \equiv \sqrt{M_W^2+(p_{\rm T}^W)^2}$.  All partons $m$ are included in
$\HTpartonicp$, whether or not they are inside jets that pass the
cuts.  This quantity is not directly measurable; however, it is very 
similar to the more usual jet-based total transverse energy, and it
is more practical for use as a dynamical scale choice.  Both $\HTpartonic$
and the modified version $\HTpartonicp$ are independent of the
experimental cuts. Thus, modifying the cuts will not affect the value
of the matrix element at a point in phase space.  This makes it
suitable as a choice of renormalization or factorization scale,
avoiding unwanted dependence on experimental cuts.  Later we will
compute the distribution in the jet-based observable
$\HTjets=\sum_{j\in {\rm jets}} p_{\rm T}^j$.  This variable is similar to the 
partonic version, $\HTpartonicp$, except that the $W$ boson $E_{\rm T}$
is omitted, and it is based on jets passing all cuts.

We define jets using the anti-$k_T$ algorithm~\cite{antikT} with
parameter $R = 0.5$.  The jets are ordered in $\pT$,
and are labeled $i,j=1,2,3,\ldots$ in
order of decreasing transverse momentum $p_{\rm T}$, with jet $1$ being the
leading (hardest) jet.  The transverse mass of the $W$ boson is
computed from the transverse momenta of its leptonic decay products,
$M_{\rm T}^W=\sqrt{2E_{\rm T}^e E_{\rm T}^\nu (1- \cos(\Delta \phi_{e\nu}))}$.

In our study, we consider the inclusive process $p p \rightarrow$
\Wjjjjj{} jets at an LHC center-of-mass energy of $\sqrt{s} = 7$ TeV with
the following set of cuts:
\begin{eqnarray}
&& E_{\rm T}^{e} > 20 \hbox{ GeV} \,, \hskip 1.5 cm 
|\eta^e| < 2.5\,, \hskip 1.5 cm 
\ETsl > 20 \hbox{ GeV}\,,  \hskip 1.5 cm \nn\\
&& \pT^\jet > 25 \hbox{ GeV}\,, \hskip 1.5 cm 
|\eta^\jet|<3\,,  \hskip 1.5 cm M_{\rm T}^W > 20  \hbox{ GeV}\,. 
\end{eqnarray}
In this study we take the missing transverse energy, $\ETsl$, to equal
the neutrino transverse energy, $E_{\rm T}^\nu$.

In carrying out the computation we imposed a set of looser cuts and
generated \root{}~\cite{ROOT} format \ntuples.  As mentioned above and
described further below, the \ntuples{} store intermediate results
such as parton momenta and coefficients associated with the event
weights for the events passing the looser cuts, whose only restriction
is that the minimum jet transverse momentum is $\pT^\jet > 25$~GeV.
The \ntuples{} are also valid for anti-$k_T$, $k_T$ and \SISCone{}
algorithms~\cite{antikT,JetAlgorithms} for $R=0.4, 0.5, 0.6, 0.7$, as
implemented in the FASTJET package~\cite{FastJet}.  In the \SISCone{}
case the merging parameter $f$ is chosen to be $0.75$.  This allows
the \ntuples{} to be used for studying the effects of varying the jet
algorithm, along with variations due to parton distributions, scale
choices, and experimental cuts.

In our study, we use the MSTW2008 LO and NLO
PDFs~\cite{MSTW2008} at the respective orders.  We use the five-flavor
running $\alpha_s(\mu)$ and the value of $\alpha_s(M_Z)$ supplied with
the parton distribution functions.

Our predictions are at parton level.  We do not apply corrections
due to non-perturbative effects such as those induced by the
underlying event or hadronization.  For comparisons to experiment it
is important to incorporate these effects, although for most cross-section
ratios we do not expect them to be large.  Parton-shower event generators
such as POWHEG and MC@NLO~\cite{POWHEGBOXMCNLO}, and further refinements of
these methods~\cite{POWHEGMCNLOfurther}, have been developed that
consistently include a parton shower and maintain NLO accuracy for
events with a specified jet multiplicity.  More recently,
advances have been made in maintaining the NLO accuracy
across different jet multiplicities in a single sample~\cite{NLOmerging}.
These advances mark an important step in significantly reducing
theoretical uncertainties associated with hadron-level predictions
of many types of LHC events.  We look forward to applying them in
the future to the production of $W$ bosons with up to five additional jets.
 
\subsection{Formalism and Software}
\label{FormalismSubsection}

The new techniques we use for obtaining virtual contributions are
collectively known as on-shell methods, and are reviewed in
refs.~\cite{OnShellReviews}.  These methods rely on underlying
properties of amplitudes --- factorization and unitarity --- in order
to express them in terms of simpler, on-shell amplitudes of lower
multiplicity.  While amplitudes necessarily contain off-shell states
inside loops or trees, avoiding direct use of these states allows the
method to avoid the gauge dependence they induce.  Eliminating
the gauge dependence greatly reduces the enormous cancellations of
intermediate terms that would plague a textbook Feynman-diagram
calculation.  The first application of the unitarity
method~\cite{UnitarityMethod} to collider physics was to obtain
the analytic matrix elements for $q\bar{q}gg \rightarrow V$ and
$q \bar{q}Q_1\bar Q_1 \rightarrow V$ ($V = W$ or $Z$)~\cite{Zqqgg},
used in the NLO program MCFM~\cite{MCFM}.  More recently, on-shell methods
been implemented in
a more flexible numerical form, breaking the long-standing bottleneck
to NLO computations for higher-multiplicity final states posed by the
one-loop (virtual) corrections.  These methods scale well as the
number of external legs
increases~\cite{OtherOnShellPrograms,GenHel,GZ,EMZW3j,W3jPRL,%
W3jDistributions,NLOttjj,NLOVVjj,W4j,Z4j}.  There
have also been important advances in computing virtual corrections
with more traditional methods~\cite{HeavyQuarkFeynman}.

One-loop amplitudes in QCD with massless quarks may be expressed as a sum
over three different types of Feynman integrals (boxes, triangles, and
bubbles) with additional so-called rational terms.  The integrals are
universal and well-tabulated, so the aim of the calculation is to compute their
coefficients, along with the rational terms.  In an on-shell approach,
the integral coefficients may be computed using four-dimensional
generalized unitarity~\cite{UnitarityMethod, Zqqgg, NewUnitarity},
while the rational terms may be computed either by a loop-level
version~\cite{GenHel} of on-shell recursion~\cite{BCFW} or using
$D$-dimensional unitarity~\cite{DDimUnitarity}.  We use a numerical
version~\cite{BlackHatI} of Forde's method~\cite{Forde} for the
integral coefficients, and subtract box and triangle
integrands similar to the Ossola--Papadopoulos--Pittau
procedure~\cite{OPP}, improving the numerical
stability.  To compute the rational terms, we use a numerical
implementation of Badger's massive continuation method~\cite{Badger},
which is related to $D$-dimensional unitarity.

These algorithms are implemented in an enhanced version of the
\BlackHat{} code~\cite{BlackHatI,W3jPRL}.  \BlackHat{} organizes the
computation of the amplitudes in terms of elementary gauge-invariant
``primitive amplitude'' building blocks~\cite{Primitive,Zqqgg}.  Many
primitive amplitudes can be associated with Feynman diagrams in which
all external partons touch the loop (i.e.~there are no nontrivial
trees attached to the loop).  Representative Feynman diagrams for the
leading-color primitive amplitudes used in the present calculation are
shown in \fig{lcdiagramsFigure}.  The primitive amplitudes are then
assembled into partial amplitudes, which are the kinematic
coefficients of the different color tensors that can appear in the
amplitude.  The complete virtual cross section is obtained by
interfering the one-loop partial amplitudes with the tree-level
amplitude and summing over spins and color indices.  The color factors
arising from the color sum in the assembly of primitive amplitudes
into partial amplitudes become highly nontrivial as the number of
quark lines increases; we use a general solution given in
ref.~\cite{ItaOzeren}.  An important feature is that each primitive
building block has a relatively simple analytic structure with only a
limited number of spurious singularities present.  A given primitive
amplitude can appear in multiple partial amplitudes and does not have
to be recomputed for each one.  This approach also allows for a
straightforward separation of leading- and subleading-color
contributions. This separation can be exploited to significantly
enhance the efficiency of the Monte Carlo
integration~\cite{W3jDistributions}: The subleading-color
contributions are much smaller, yet more computationally costly;
separating them out allows them to be evaluated at far fewer
phase-space points than the leading-color contributions, in order to
obtain similar absolute uncertainties.

In the \Wjjjjx-jet calculations we drop altogether the small
but time-consuming subleading-color contributions to the virtual
corrections.  As
explicitly verified for three-~\cite{W3jDistributions} and four-jet
production~\cite{ItaOzeren}, the omitted
subleading-color contributions to the virtual corrections are
typically 10\% of the leading-color virtual terms, and under 3\% of the
total cross section.  We expect the dropped subleading-color contributions
to be similarly small for \Wjjjjj-jet production.  The precise version
of the leading-color approximation for the virtual terms used here is
the one of ref.~\cite{ItaOzeren}.  It retains
full-color dependence in all contributions multiplying poles in the
dimensional regularization parameter $\epsilon$ in the virtual
corrections.  In the finite parts of the virtual corrections,
it drops certain contributions that are subleading in the number of
colors $N_c$ in the formal limit $N_c\rightarrow \infty$, with
$n_{\it f}/N_c$ held fixed.  In particular, we drop those finite parts of the
leading-color partial amplitudes that are suppressed by explicit powers
of $1/N_c$, as well as all finite parts of the subleading-color partial
amplitudes.  In forming the color-summed interference of the surviving
parts of the one-loop amplitudes with the tree amplitudes, we do not drop
any further terms.
The \BlackHat{} code uses the four-dimensional helicity (FDH)
scheme~\cite{FDH} internally but automatically shifts
the result before output to the 't~Hooft--Veltman scheme~\cite{HV}
using the full-color dependence in the shift.  The approximation
differs in these last two aspects from the one used in the earlier
\BlackHat{} calculation of \Wjjjj-jet production~\cite{W4j}.  This
causes very slight shifts, under a percent, in our reported
cross sections for \Wjn{}-jet production for $n\le 4$ compared to
ref.~\cite{W4j}.  (A somewhat larger shift arises from the choice of a
five-flavor scheme for the running of the coupling instead of the
six-flavor scheme used earlier.)

The NLO result also requires real-emission corrections to the LO
process, which arise from tree-level amplitudes with one additional
parton; sample contributions are illustrated in
\fig{TreeDiagramsFigure}.  For $W$ production with 
four or five associated jets we use
the~\COMIX{} code~\cite{Comix}, included in the \SHERPA{}
framework~\cite{Sherpa}, to compute these contributions, including the
Catani-Seymour dipole subtraction terms~\cite{CS}.  The \COMIX{} code
is based on a color-dressed form~\cite{CDBG} of the Berends-Giele
recursion relations~\cite{BG}, making it very efficient for processes
with high multiplicities.  When there are fewer jets, we use the
\AMEGIC{} package~\cite{Amegic} instead. In order to carry out the
Monte Carlo integration over phase space we use an efficient
hierarchical phase-space generator based on QCD antenna
structures~\cite{AntennaIntegrator}, as incorporated into \SHERPA.
The integration is performed using an adaptive
algorithm~\cite{Vegas}. 
Running the algorithm in a stable and convergent manner
for the real-emission contributions is highly nontrivial, in particular
given the intricate structure of their infrared subtractions.

In general, the same physical distributions need to be analyzed at
different PDF error sets, different renormalization or factorization
scales, and for different jet algorithms or experimental cuts.
We have organized the computation so
the matrix elements do not have to be reevaluated for each choice of
parameters~\cite{FourJets}.  For each event we generate, we record the
momenta for all partons, along with the numerical values of the
coefficients of the various
scale- or PDF-dependent functions.  Each term contains a simple
function we wish to vary, such as a logarithm of the renormalization
scale, multiplied by a numerical coefficient independent of such
variation.  We store the intermediate information in \root{}-format
\ntuple{} files~\cite{ROOT}.  At the end of the main computation
we assemble the stored matrix-element coefficients, the PDF and scale
choices to obtain cross sections.  The availability of these
intermediate results makes it
straightforward to evaluate cross sections flexibly, 
for different scales, PDF error sets, experimental cuts or jet-based
observables.  This format has also been
used by the experimental collaborations to compare results 
from \BlackHat{} $+$ \SHERPA{} to experimental data~\cite{NTupleUse}.

\subsection{Numerical Stability}

\begin{figure}[t]
\hskip -12 mm 
\includegraphics[clip,scale=0.32]{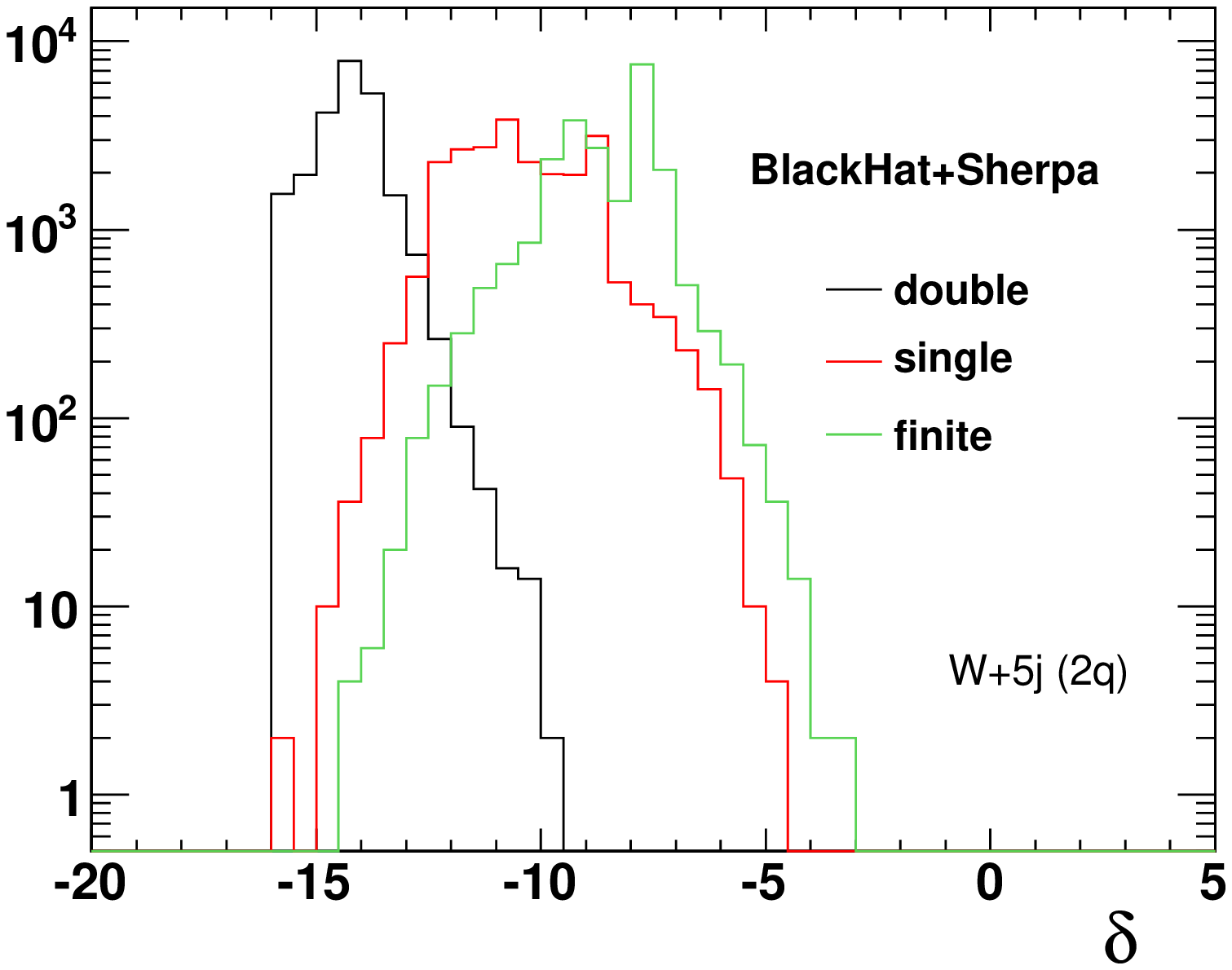} \hskip -10 mm 
\includegraphics[clip,scale=0.32]{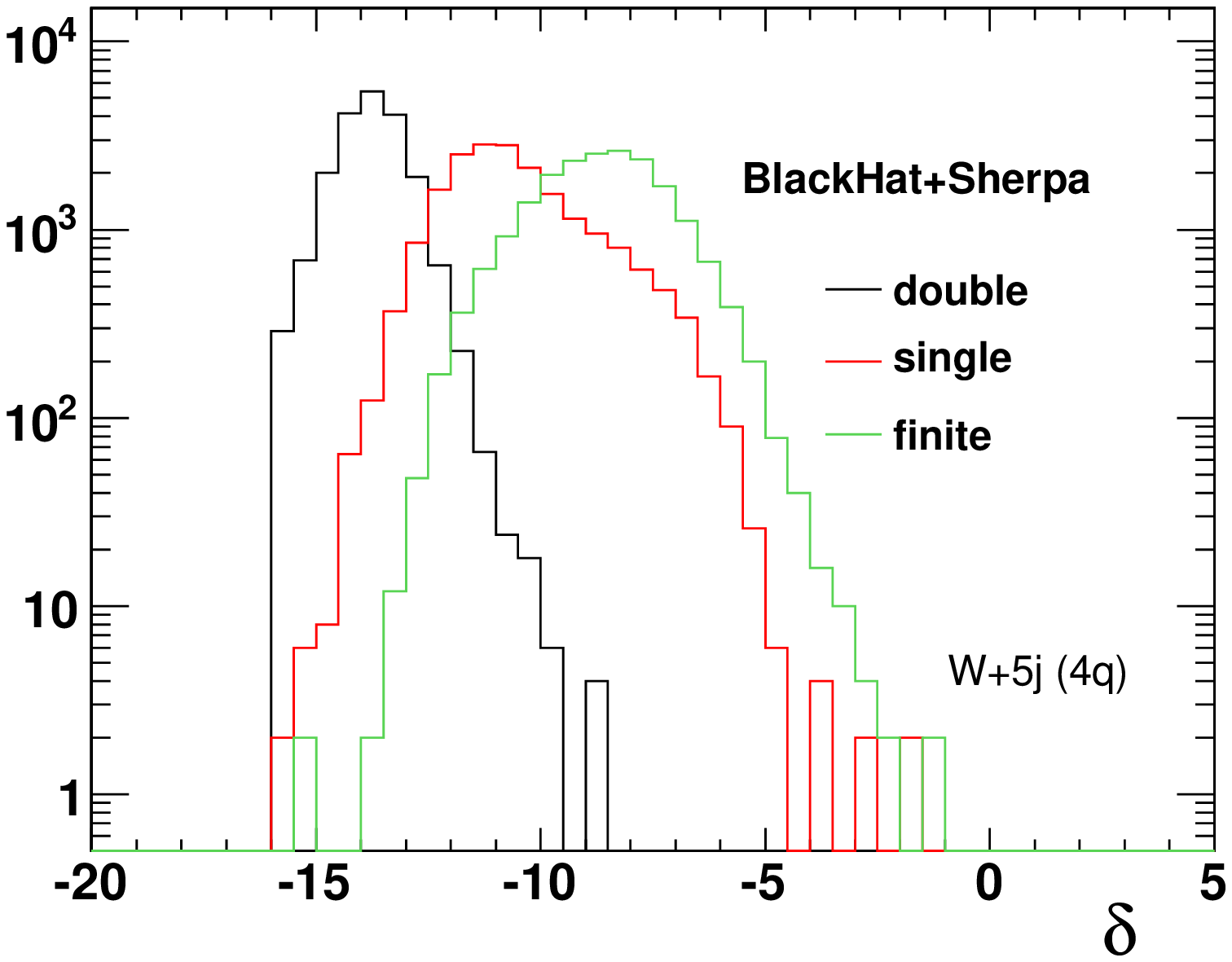} \hskip -10 mm 
\includegraphics[clip,scale=0.32]{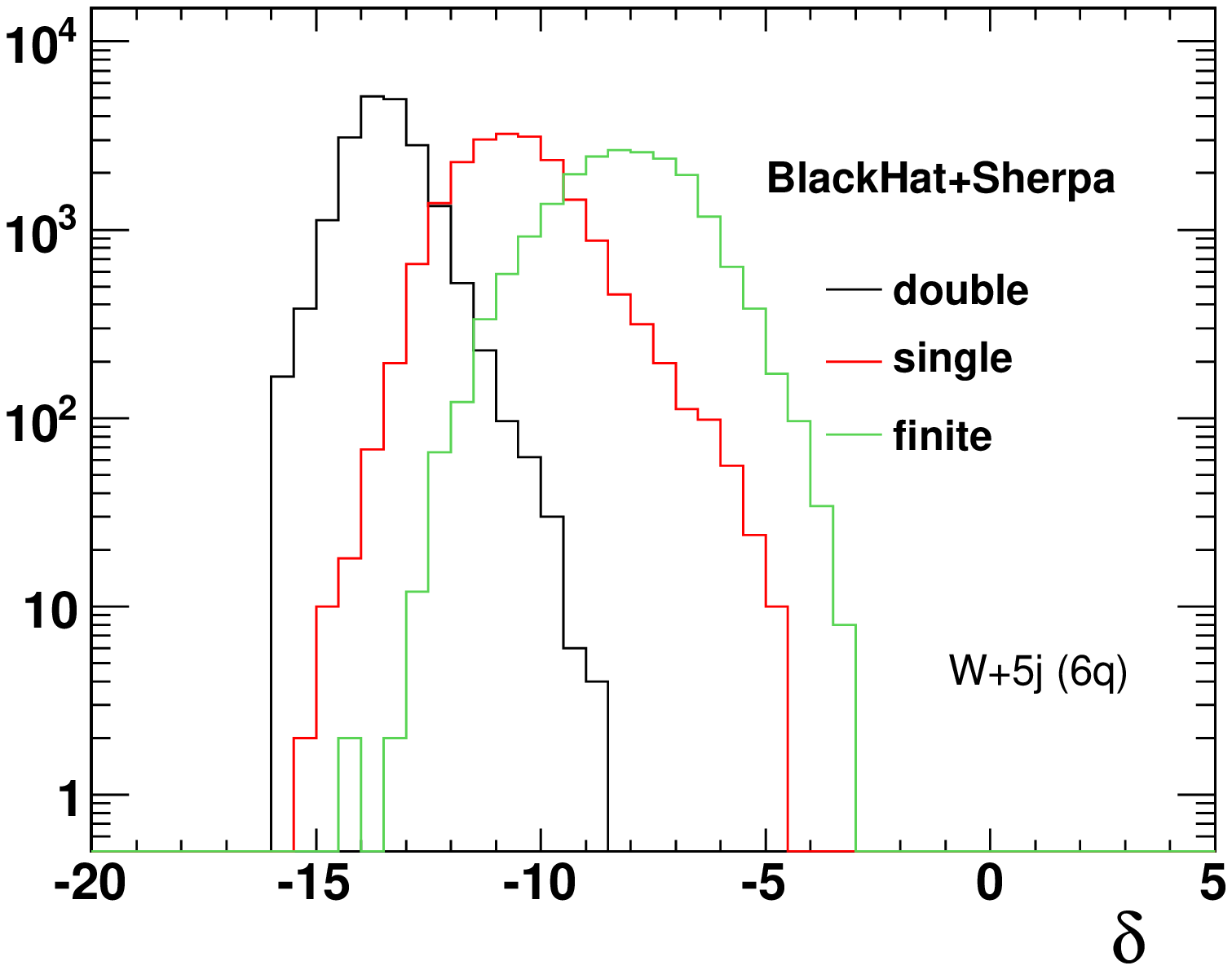} \hskip -1.1 cm $\null$
\caption{ The distribution of the relative error in the virtual cross
  section for three subprocesses, $g d\rightarrow e^- \bar\nu_e gggg u$,
  $u d\rightarrow e^- \bar\nu_e ggg uu$ and
  $ud \rightarrow e^- \bar \nu_e g u u s \bar s$, reading from left
  to right.  The horizontal axis is the logarithm of the relative
  error~(\ref{RelError}) between an evaluation by \BlackHat{}, running
  in production mode, and a target expression evaluated at higher
  precision.  The vertical axis shows the number of phase-space points
  having that relative error.  The dark (black) line
  labeled by ``double'' shows the $1/\eps^2$ term; the darkly shaded
  (red) curve labeled by ``single'', the $1/\eps$ term; and the
  lightly shaded (green) curve labeled by ``finite'', the finite
  ($\eps^0$) term.  Each plot is based on approximately 10,000 points
  in phase space, distributed as in the actual calculation.
  }
\label{StabilityFigure}
\end{figure}

The different terms in the virtual contributions to matrix elements typically
contain poles at unphysical locations in phase space.  These poles cancel
out when summing over all terms.  When different terms are computed
independently, one must ensure that the numerical precision of the
computations suffices for cancellation
of these spurious singularities anywhere in the phase space, so as to avoid unwanted
loss of precision in the full matrix element.  While the degree of the spurious
singularities is in fact typically lower when using on-shell methods than with
traditional ones, they are nonetheless present.  They may cancel between coefficients
of different integrals, each computed numerically.  
Obtaining a numerically stable result for the virtual terms at each point
in phase space is accordingly nontrivial.  

 The \BlackHat{} code detects instabilities
following the criteria described in refs.~\cite{BlackHatI,W3jDistributions}.
When faced with an unstable point in phase space, the code switches to higher
precision arithmetic, and recomputes only those terms which suffer from the
instability.  The higher-precision computations are performed in software
rather than in hardware, and are accordingly much slower than those at native
precision.  The recomputation of a limited number of terms (as opposed to the
entire amplitude) minimizes the additional computer time.  We use the {\sc QD}
package~\cite{QD} for higher-precision arithmetic.

In \fig{StabilityFigure}, we illustrate the stability of the
virtual contribution to the differential cross section, $d\sigma_V$, summed
over colors and over all helicity configurations for the three
subprocesses,
$g d\rightarrow e^- \bar\nu_e gggg u$
and $u d\rightarrow e^- \bar\nu_e ggg uu$ and 
$u d\rightarrow e^- \bar\nu_e g u u s \bar s$.
In each plot, the horizontal axis represents the logarithmic error,
\begin{equation}
\delta = \log_{10}\left(\frac{|d \sigma_V^{\rm BH}- d \sigma_V^{\rm target}|}
          {| d \sigma_V^{\rm target}|} \right),
\label{RelError}
\end{equation}
for each of the three components: $1/\epsilon^2$, $1/\epsilon$ and
$\epsilon^0$, where $\eps = (4-D)/2$ is the dimensional regulator.  In
\eqn{RelError}, $d\sigma_V^{\rm BH}$ is the cross section computed by
\BlackHat{} as it normally operates.  The target value $d\sigma_V^{\rm
  target}$ is the cross section computed by \BlackHat{} using
multiprecision arithmetic with approximately $32$ digits, and
approximately $64$ digits if the point is deemed unstable using the
criteria described in refs.~\cite{BlackHatI,W3jDistributions}.  The
phase-space points are selected in the same way as those used to
compute cross sections.  We note that an overwhelming majority of
events are accurate to better than one part in $10^3$ --- that is, to
the left of the `$-3$' mark on the horizontal axis.  We have
explicitly checked that the few points to the right of this mark
produce completely negligible errors in the final cross section or
distributions, as their cross-section values are not especially large.


\section{Results}
\label{ResultsSection}

We now present our NLO results for \Wjjjjj-jet production at the LHC.
We first discuss the renormalization-scale dependence of the total
cross section.  Then we provide the total hadronic energy distribution
as an example distribution.  Finally we present results for the total
cross sections for \Wmjjjjj-jet and \Wpjjjjj-jet production and for
the $\pT$ distributions of the five jets.

\subsection{Scale dependence}

\begin{figure}[t]
\includegraphics[clip,scale=0.65]{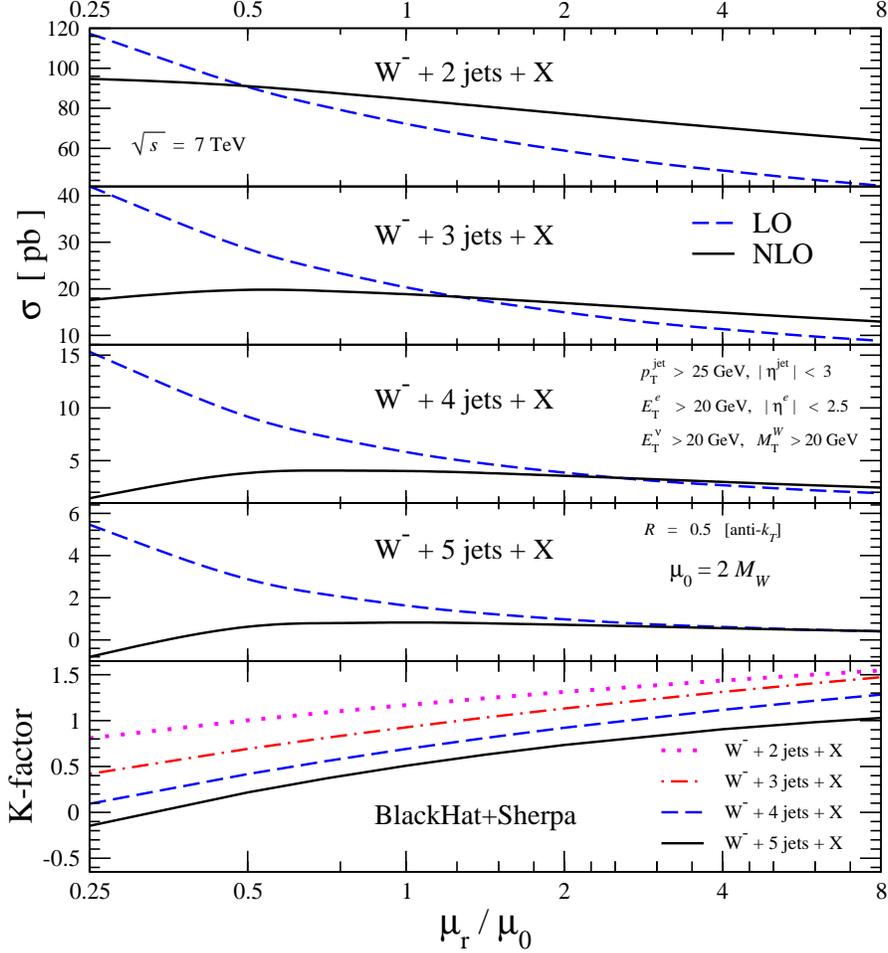}
\caption{ 
 The renormalization-scale dependence of the total cross
  section using a fixed reference scale of $\mu_0 = 2 M_W$.  The top four
  panels give the renormalization-scale dependence at both LO and NLO for
  \Wjj-jets through \Wjjjjj-jets.  The bottom panel shows the $K$
  factors for these cases, with the top curve for \Wjj-jets and the
  bottom one for \Wjjjjj-jets. The factorization scale is held fixed.
}
\label{ScalesTotXSFigure}
\end{figure}

\begin{figure}[t]
\includegraphics[clip,scale=0.45]{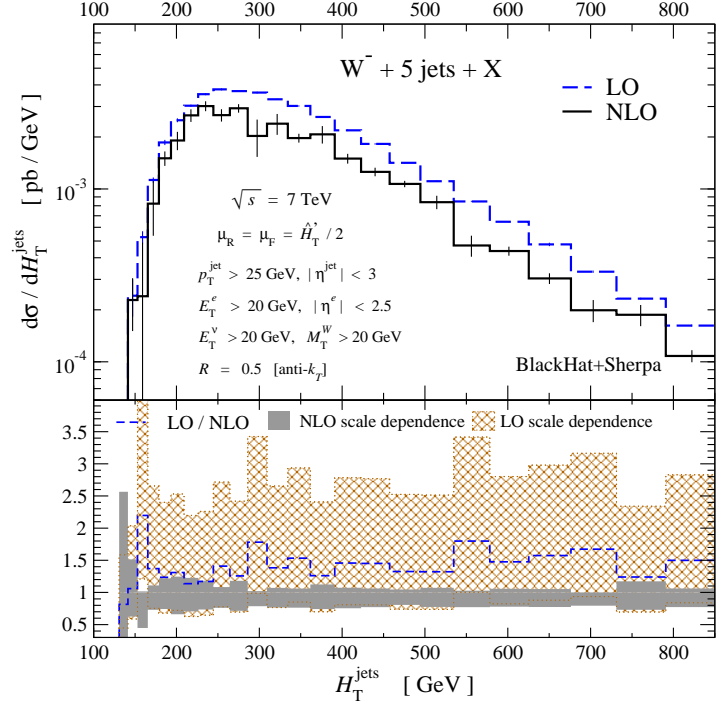}
\caption{The improvement in the renormalization and factorization
  scale dependence of the differential cross section as a function of
  the hadronic total transverse energy $H_{\rm T}^{\rm jets}$,
  comparing LO to NLO at the LHC at $\sqrt{s}=7$~TeV.  In the upper
  panels, the NLO predictions are shown as solid (black) lines, while
  the LO predictions are shown as dashed (blue) lines.  The thin
  vertical line in the center of each bin (where visible) gives its
  numerical integration error, corresponding to the fluctuations in
  the plots. The lower panels show the predictions for the LO
  distribution and scale-dependence bands, normalized to the NLO
  prediction at the scale $\mu=\HTpartonic'/2$.  The LO distribution
  is the dashed (blue) line, and the scale-dependence bands are shaded
  (gray) for NLO and cross-hatched (brown) for LO.  }
\label{scalesHTFigure}
\end{figure}

We expect perturbative results to be more stable under variation of the
renormalization and factorization scales as the perturbative order
is increased.  The residual variability has been used as a
proxy for the expected uncertainty due to higher-order corrections
beyond the calculated order.  In previous
papers~\cite{W3jDistributions,BlackHatZ3jet}, we have seen that the
variability increases substantially with a growing number of jets at LO,
but stabilizes at under 20\% at NLO.  This trend continues as the number
of jets grows beyond the multiplicity considered in our just-cited studies.
In \fig{ScalesTotXSFigure}, we show the variation of the total cross
section for \Wmjn-jet production with the renormalization scale
around a central choice of $\mu_0=2 M_W$, for $n=2,3,4,5$ at LO and at
NLO, along with the so-called ``$K$ factor,'' the ratio of the NLO to LO
cross sections.  We vary the scale down by a factor of four and
upwards by a factor of eight.  A fixed scale of
$\Ord(M_W)$ is appropriate for the total cross section, as it is
dominated by total transverse energies of the order of a small
multiple of this scale.  We hold the factorization scale fixed
in order to eliminate changes in the PDFs as we vary the scale.
This makes it simpler to see the trends as we change from two to five jets.
Similar improvements in scale dependence are also observed when we
include the variation of the factorization scale.  

The four upper panels of \fig{ScalesTotXSFigure} show that the scale
variation at NLO is greatly reduced with respect to that at LO.
Furthermore, the LO variation grows substantially with an increasing
number of jets, while the NLO variation is fairly stable.  This
increase is expected, because there is an additional power of
$\alpha_s$ for every additional jet; the variation of $\alpha_s$ is
uncompensated at LO, but compensated at NLO by the virtual
corrections.  The relative stability of the NLO prediction in contrast
to the LO one is also reflected in the bottom panel.  As one increases
the number of jets from the smallest (two), the $K$-factor curve
steepens at first.  This steepening slows down as the number of jets
reaches five.  This reflects a slowing down of the relative
stabilization with growing number of jets.  For the \Wjjjjj-jet process at
LO, the change in the total cross section is on the order of a factor
of $2$ if we vary the renormalization scale by a factor of 2 around
$M_W$ as in \fig{ScalesTotXSFigure}.  In contrast, at NLO the
dependence is cut to about $\pm20$\%.

The distributions we study have a large dynamic range.  Accordingly,
for physics studies we choose an event-by-event scale to match typical
energy scales individually rather than merely on average.  Following
ref.~\cite{W3jDistributions}, we use a central scale equal to
half the total partonic final-state transverse energy,
\begin{equation}
\mu_R = \mu_F = \mu = \HTpartonic'/2\,.
\label{ScaleChoice}
\end{equation}
where $ \HTpartonic'$ is defined in \eqn{HTp}.
As an illustration of the scale dependence in a distribution using
this choice, we show the variation of the LO and NLO \Wjjjjj-jet
cross section as a function of the total jet transverse energy
$\HTjets$ in \fig{scalesHTFigure}.  The bands in the figure
show the results from varying the scale up and down by a factor of 2
around the central value (\ref{ScaleChoice}), taking
the minimum and maximum of the observable evaluated at five
values: $\mu/2, \mu/\sqrt{2}, \mu, \sqrt{2}\mu, 2\mu$.  The
figure shows the markedly reduced scale dependence at NLO compared to
that at LO. It also shows a remarkably
flat ratio between the LO and NLO distributions. Other authors have suggested
alternate choices of dynamical scale~\cite{OtherScaleChoices,BDDP}.


\begin{table}
\vskip .4 cm
\centering
\begin{tabular}{||c||c|c||c|c|}
\hline
Jets &  $W^-$ LO  & $W^-$ NLO & $W^+$ LO & $W^+$ NLO   \\  \hline

1 & $284.0(0.1)^{+26.2}_{-24.6}$ & $351.2 (0.9)^{+16.8}_{-14.0}$ &
  $416.8(0.6)^{+38.0}_{-35.5}$ & $516(3)^{+29}_{-23}$\\ \hline
2 & $83.76(0.09)^{+25.45}_{-18.20}$ & $83.5(0.3)^{+1.6}_{-5.2}$ & 
$130.0(0.1)^{+39.3}_{-28.1}$ & $125.1(0.8)^{+1.8}_{-7.4}$\\ \hline
3 & $21.03(0.03)^{+10.66}_{-6.55}$ & $18.3(0.1)^{+0.3}_{-1.8}$ &
 $34.72(0.05)^{+17.44}_{-10.75}$ & $29.5(0.2)^{+0.4}_{-2.8}$\\ \hline
4 & $4.93(0.02)^{+3.49}_{-1.90}$ & $3.87(0.06)^{+0.14}_{-0.62}$ &
 $8.65(0.01)^{+6.06}_{-3.31}$ & $6.63(0.07)^{+0.21}_{-1.03}$\\ \hline
5 & $1.076(0.003)^{+0.985}_{-0.480}$ & 
$0.77(0.02)^{+0.07}_{-0.19}$ & $2.005(0.006)^{+1.815}_{-0.888}$ &
 $1.45(0.04)^{+0.12}_{-0.34}$\\ \hline
\end{tabular}
\caption{Total cross sections in pb for \Wjn{} jet production at the
LHC at $\sqrt{s}=7$~TeV, using the  anti-$k_T$ jet algorithm with  $R=0.5$.
The NLO results for \Wjjjjx{}-jet production use the leading-color
approximation discussed in the text.  The numerical
  integration uncertainty is given in parentheses, and the scale dependence
  is quoted in superscripts and subscripts.
}
\label{CrossSectionAnti-kt-R5Table}
\end{table}

\begin{figure*}[t]

\includegraphics[clip,scale=0.64]{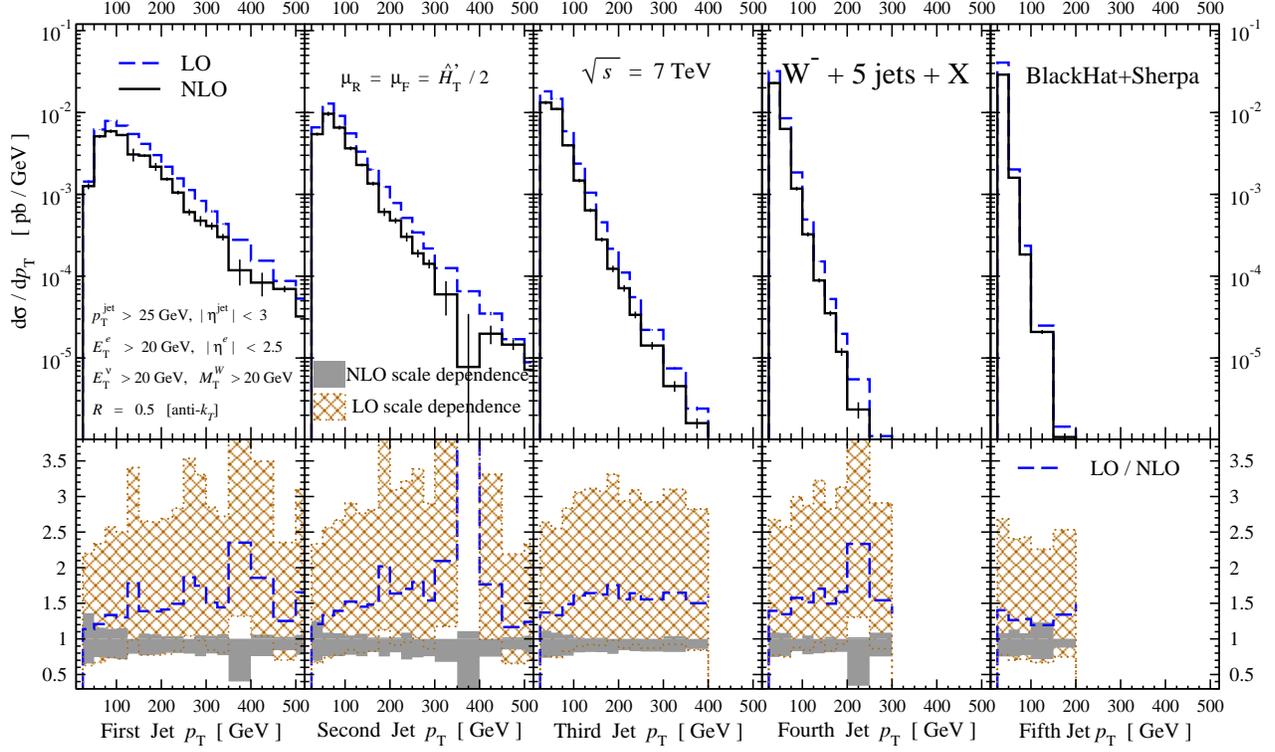}
\caption{The $\pT$ distributions of the leading five jets in
  \Wmjjjjj{}-jet production at the LHC at $\sqrt{s}=7$~TeV.  In the
  upper panels, the NLO predictions are shown as solid (black) lines,
  while the LO predictions are shown as dashed (blue) lines.
  The lower panels show the
  predictions for the LO distribution and scale-dependence bands
  normalized to the NLO prediction (at the scale $\mu=\HTpartonic'/2$).
The LO distribution is the dashed (blue) line, and the 
scale-dependence bands are shaded (gray) for NLO and cross-hatched
  (brown) for LO.  
}
\label{Wm5ptFigure}
\end{figure*}

\begin{figure*}[t]
\includegraphics[clip,scale=0.64]{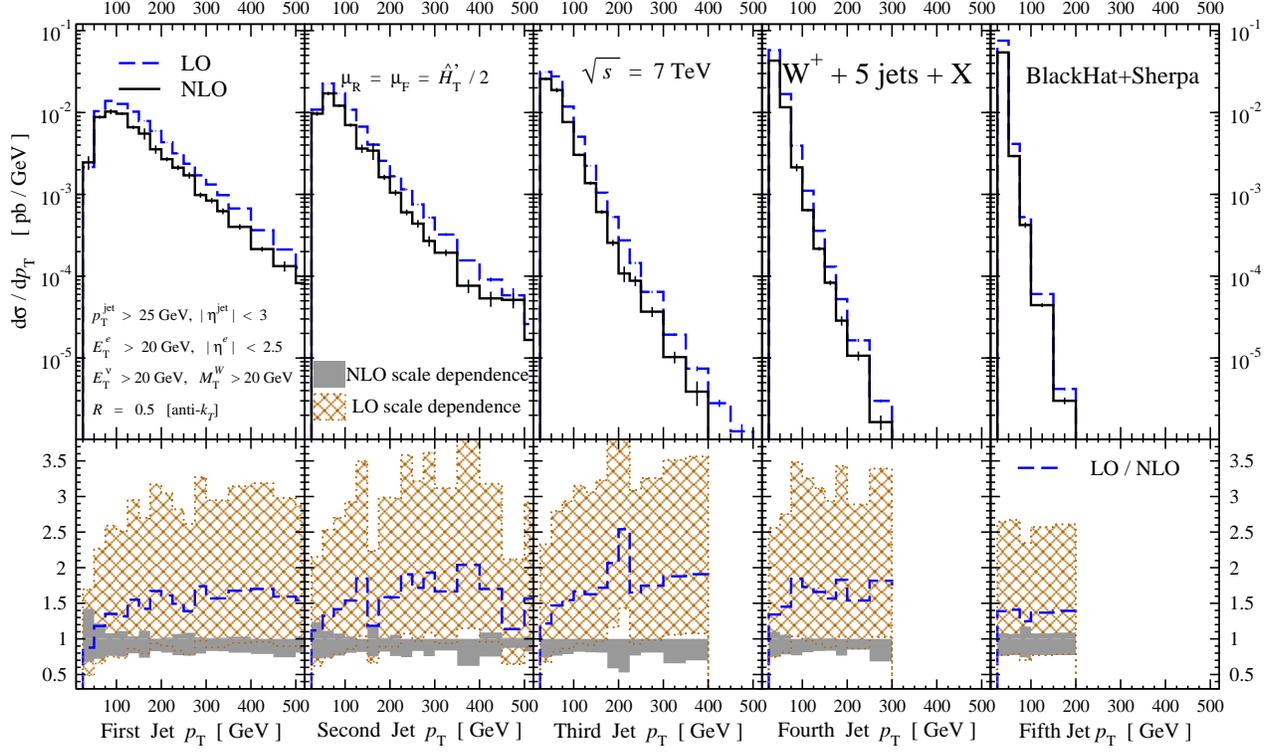}
\caption{The  $\pT$ distributions of the leading five jets in
  \Wpjjjjj{}-jet production at the LHC at $\sqrt{s}=7$~TeV.
}
\label{Wp5ptFigure}
\end{figure*}

\subsection{Cross Sections and Distributions}
\label{CrossSectionsSection}

In~\Tab{CrossSectionAnti-kt-R5Table}, we present the LO and NLO
parton-level cross sections for inclusive $W^-$- and $W^+$-boson
production accompanied by one through five jets.  As discussed in
\sect{BasicSetUpSection}, we include all subprocesses, except for the
IR-subtracted real-emission contributions with four quark pairs, which
give contributions (as determined from a low-statistics evaluation)
of well under 1\% using \SHERPA{}'s $\alpha_{\rm dipole} = 0.03$.  
Neglecting these contributions leaves a residual dependence on
$\alpha_{\rm dipole}$; however, it is numerically unimportant to the full
result for $\alpha_{\rm dipole} = 0.03$.  We perform
a full-color sum everywhere, except in the virtual
contributions to \Wjjjjx-jet production.  In these latter
contributions, we employ the leading-color approximation discussed
in~\sect{FormalismSubsection},
which has been validated to be accurate to better than 3\% for
$W$-boson production in association with up to four
jets~\cite{W3jDistributions,ItaOzeren}.

In \figs{Wm5ptFigure}{Wp5ptFigure}, we show the $\pT$ distributions
for the five leading jets in \Wmjjjjj-jet and \Wpjjjjj-jet production
at $\sqrt{s}=7$~TeV at the LHC.  In the upper panels, we show the
distributions at LO and NLO on a logarithmic scale.  On this scale,
the differences between distributions are not easily seen, so we
display the ratios to the NLO prediction (with the central scale
choice $\mu = \HTpartonic'/2$) in the lower panels.  We also show the
scale-dependence bands for both the LO and NLO predictions, again
generated by varying the scale up and down by a factor of 2. We can
see that the scale dependence is dramatically smaller at NLO,
fulfilling one of the goals of the calculation.  It makes the
prediction of this high-multiplicity process truly quantitative.
At larger transverse momenta for the leading two jets, the bands do show
noticeable fluctuations because of limited statistics.

The overall normalization is not the only feature that changes
in going from LO to NLO.  The shape of
the last-jet distribution appears to be the same at NLO (up to 
fluctuations from limited integration statistics), but all harder
jets --- four, in the present study --- appear to have slightly softer
distributions at NLO compared to LO.  This continues a
pattern seen previously in \Wjjj-jet~\cite{W3jPRL,W3jDistributions} 
and \Wjjjj-jet~\cite{W4j} production.

\begin{table}
\centering
\begin{tabular}{||c||c|c||c|c||c|c||}
\hline
\multirow{2}{*}{Jets} &\multicolumn{2}{c||}{$W^+/W^-$} &
\multicolumn{2}{c||}{$\displaystyle\vphantom{\sum_{i=1}^W}\frac{W^-+ n}{W^-+(n\!-\!1)}$}&
\multicolumn{2}{c||}{$\displaystyle\vphantom{\sum_{i=1}^W}\frac{W^+ +n}{W^+ +(n\!-\!1)}$}\\
\cline{2-7}
& LO  &  NLO & LO & NLO & LO & NLO \\  \hline
1 & $\; 1.467(0.002)\; $ & $\; 1.47(0.01)\; $ & --- & --- & --- &---  \\
2 & $\; 1.552(0.002)\;$ & $\; 1.50(0.01)\; $ & $\; 0.2949(0.0003)\; $ &
    $\;0.238(0.001)\;$ &  $\; 0.3119(0.0005)\;$ & $\; 0.242(0.002)\; $  \\
3 & $\; 1.651(0.003)\;$ & $\; 1.61(0.01)\; $ & $\; 0.2511(0.0005)\; $ & 
    $\; 0.220(0.001) \;$ &  $\; 0.2671(0.0004)\; $ & $\; 0.235(0.002)\;$  \\
4 & $\; 1.753(0.006)\;$ & $\; 1.72(0.03)\; $ & $\; 0.2345(0.0008)\; $ &
     $\; 0.211(0.003)\; $ &  $\; 0.2490(0.0005)\;$ & $\; 0.225(0.003)\; $  \\
5 & $\; 1.864(0.008)\;$ & $\; 1.87(0.06)$ & $ 0.218(0.001)\;\;\; $ & 
     $\; 0.200(0.006)\; $ &  $\; 0.2319(0.0008)\;$ & $\; 0.218(0.006)\; $  \\
\hline
\end{tabular}
\caption{The first two columns give cross-section ratios for $W^+$
  production to $W^-$ production, as a function of the number of associated
 jest.  The last two columns give the ratios of the cross section for 
 the given process to that with one
  fewer jet.  The numerical integration uncertainty is in parentheses.
 }
\label{CrossSectionRatiosTable}
\end{table}

In \Tab{CrossSectionRatiosTable}, we present the charge-asymmetry
ratio, which is the ratio between the production of a $W^+$ boson
to a $W^-$ boson, each accompanied by
up to five jets. This ratio can be a sensitive probe of new
physics~\cite{KomStirling}.  The table also shows the
jet-production ratios~\cite{JetProductionRatio,CMSWjets} for either
sign of the $W$ charge, here defined by the
ratio of the total cross section for \Wpmjn-jet to \Wpmjnm-jet
production.  The charge-asymmetry ratios are all significantly greater
than unity, and grow with increasing numbers of jets.   The jet-production
ratios are of order $1/4$, and decrease with increasing numbers of jets.
The NLO corrections to the charge-asymmetry are quite small, and the 
corrections to the jet-production ratios are modest but noticeable.

These values of the charge-asymmetry ratio reflect the excess of up
quarks over down quarks in the proton.  The $W^+$ bosons are
necessarily emitted by up-type quarks, whereas $W^-$ bosons are
emitted by down-type quarks.  The up-quark excess in the proton then
leads to larger $W^+$ cross sections.  As the number of jets
increases, production of a $W$ requires a larger value of the momentum
fraction $x$.  This alters the mix of subprocesses that contribute to
vector-boson production, and also increases the $u(x)/d(x)$ ratio.
The case of \Wj-jet production is special, because the $gg$ initial
state is absent at LO.  In general, the $gg$ initial state contribution is expected
to decrease with increasing $x$; but for $W^−$ production this contribution actually increases with the number of jets, for production in association with two through four jets (judged by LO fractions); only for more than four jets does it start to decrease as expected. (In $W^+$ production, it starts to decrease above three rather than four
accompanying jets.)  The $qg$ initial state decreases, but the $qq$
initial state increases; the net effect, along with the increase in
the $u(x)/d(x)$ fraction, is an increase in the $W^+/W^-$ ratio.  The
results presented here extend our previous NLO analysis of these
ratios with up to four accompanying jets~\cite{W4j} to the case of
five accompanying jets.  Both the LO and the NLO
ratios are quite insensitive to correlated variations of the
renormalization and factorization scales in numerator and denominator,
so we do not quote the variation here, and only show the uncertainty
from the numerical integration.

The increase in typical values of $x$ with increasing numbers of jets also
reduces the values of both the strong coupling $\alpha_s$ and the derivatives
of the parton distributions, leading to a decrease in the jet-production
ratios.  The double ratios --- the ratios of the LO cross sections
in the fifth column of \tab{CrossSectionRatiosTable} to those in the third column, and of the
NLO cross sections in the last column to those in the fourth column --- are roughly constant,
suggesting that the decrease is primarily due to the decrease of $\alpha_s$
with increasing scale.

We can use the ratios to extrapolate to larger number of jets for the
cuts used in this study.  This approach was recently investigated 
using jet calculus and found to be a good approximation when
the jets are required to have the same minimum transverse
momenta~\cite{Gerwick}.
(Substantially different cuts could give
rise to different behavior.) While ratios involving \Wj-jet production
behave differently from the rest, both because of strong kinematic
constraints and because of missing production channels (at LO), the
remaining ratios turn out to allow an excellent fit to a straight
line.  For the charge ratio (\Wpjn{} to \Wmjn), the results through
\Wjjjj-jet production would suffice to yield a non-trivial prediction;
the ratio for \Wjjjjj-jet production confirms this prediction.  For
the jet-production ratios (\Wpmjn{} to \Wpmjnm), the \Wjjjjj- to
\Wjjjj-jet production ratio is essential to making a non-trivial
prediction, as it provides a third point in the fit.  While these
extrapolations should not be taken to too large a number of jets $n$,
we expect them to provide a reasonable prediction for $n$ somewhat
beyond five.

For the LO charge ratio with $n$ jets, we obtain the following prediction 
($n\ge 2$),
\begin{equation}
R_{W^+\!/W^-}^{\rm LO} = 1.347 \pm 0.006+(0.102 \pm 0.002)\, n\,;
\end{equation}
for the NLO ratio,
\begin{equation}
R_{W^+\!/W^-}^{\rm NLO} = 1.27 \pm 0.03+(0.11 \pm 0.01)\, n\,.
\label{WpWmNLOfit}
\end{equation}

For the $W^-$ jet-production ratio at LO, we find the following prediction ($n\ge 3$),
\begin{equation}
R^{{\rm LO,\;}W^-}_{n/(n-1)} = 0.301 \pm 0.002-(0.0165 \pm 0.0005)\, n\,;
\end{equation}
at NLO, we find,
\begin{equation}
R^{{\rm NLO,\;}W^-}_{n/(n-1)} = 0.248 \pm 0.008-(0.009 \pm 0.002)\, n\, .
\label{WminusNLOfit}
\end{equation}
Similarly, for the $W^+$ jet-production ratio at LO, we find the
following prediction ($n\ge 3$),
\begin{equation} 
R^{{\rm LO,\;}W^+}_{n/(n-1)} = 0.320 \pm 0.002-(0.0177 \pm 0.0004)\, n\,;
\end{equation}
and at NLO, we find,
\begin{equation}
R^{{\rm NLO,\;}W^+}_{n/(n-1)} = 0.263 \pm 0.009-(0.009 \pm 0.003)\, n\,.
\label{WplusNLOfit}
\end{equation} 
The slopes for $W^+$ and $W^-$ differ slightly at LO but are essentially the
same at NLO.  These predictions are based on fits to the data in
\tab{CrossSectionRatiosTable}. (More precisely, they are based on an
ensemble of 10,000 fits to synthetic data distributed in a Gaussian
according to the cross sections and statistical error in
\tab{CrossSectionAnti-kt-R5Table} from which the ratios in
\tab{CrossSectionRatiosTable} were computed.)

From \eqns{WminusNLOfit}{WplusNLOfit}, we obtain the following predictions
for the NLO cross sections for production of a $W^\pm$ in association with
six jets,
\begin{eqnarray}
&&W^- + 6 \hbox{ jets}:\quad 0.15 \pm 0.01\ \hbox{pb}\,,\nonumber\\
&&W^+ + 6 \hbox{ jets}:\quad 0.30 \pm 0.03\ \hbox{pb}\,,
\end{eqnarray}
matching the experimental cuts used for \tab{CrossSectionAnti-kt-R5Table}.
The ratio of these two predictions, $2.0 \pm 0.3$,
is consistent with an extrapolation using \eqn{WpWmNLOfit},
$1.94 \pm 0.08$.

\section{Conclusions}
\label{ConclusionSection}

In this paper, we presented the first NLO QCD results for inclusive
\Wjjjjj-jet production at the LHC at $\sqrt{s}=7$~TeV.
This process is an important background to many new physics
searches involving missing energy, as well as to precise top-quark
measurements.  In addition to its phenomenological usefulness, it also
sets a new bar for the state of the art in perturbative QCD at
next-to-leading order, at hadron-collider processes with six
final-state objects including jets.  

We have adopted a number of approximations in this work: the
leading-color approximation for the virtual terms; neglecting
top-quark loops; and neglecting infrared-finite parts of real-emission
contributions with four quark pairs.  Based on studies of \Wjjjj-jet
production and $W$ production in associated with fewer
jets~\cite{W3jDistributions,ItaOzeren}, we expect the leading-color
approximation to change cross sections by no more than 3\%, and the
other approximations to be smaller yet.  Hence these approximations
should have no phenomenological significance, given the other theoretical
uncertainties.

We find a dramatic reduction in scale dependence in NLO predictions
for the total cross section, and for differential distributions as
well.  The scale dependence of observables shrinks from more than a factor of
two variation at LO to a 20\% sensitivity at NLO for \Wjjjjj-jet
production.  With our dynamical scale choice in \eqn{ScaleChoice},
we find $K$ factors typically between 0.6 and 1,
with moderate though non-trivial changes in shapes of distributions.

We have studied a number of ratios, between $W^+$ and $W^-$
production, and for processes differing by the addition of one jet.
The QCD corrections to these ratios are more modest than to total
cross sections, and they should also benefit from milder experimental
systematical uncertainties.  The ratios show interesting trends with
increasing number of jets, and with results for \Wjjjjj-jet production
in hand, we can make plausible extrapolations to results for
additional jets.  These ratios also probe the evolution of different
subprocesses with increasing parton fraction $x$.

The present study brings an unprecedented level of precision to
\Wjjjjj-jet production.  We look forward to comparing the NLO results
for this process, and for the extrapolations to yet higher numbers of jets
based upon it, with LHC data.

We thank Joey Huston, David Saltzberg, Maria Spiropulu, Gerben
Stavenga, Eric Takasugi and Matthias Webber for helpful discussions.
This research was supported by the US Department of Energy under
contracts DE--FG03--91ER40662 and DE--AC02--76SF00515.  DAK's research
is supported by the European Research Council under Advanced
Investigator Grant ERC--AdG--228301.  DM's work was supported by the
Research Executive Agency (REA) of the European Union under the Grant
Agreement number PITN--GA--2010--264564 (LHCPhenoNet). The work of KJO
and SH was partly supported by a grant from the US LHC Theory
Initiative through NSF contract PHY--0705682.  This research used
resources of Academic Technology Services at UCLA, and of the National
Energy Research Scientific Computing Center, which is supported by the
Office of Science of the U.S. Department of Energy under Contract
No.~DE--AC02--05CH11231.


\end{document}